\documentclass[preprint]{aastex}

\received{}
\accepted{}

\def\gsim{\;\rlap{\lower 2.5pt
 \hbox{$\sim$}}\raise 1.5pt\hbox{$>$}\;}
\def\lsim{\;\rlap{\lower 2.5pt
   \hbox{$\sim$}}\raise 1.5pt\hbox{$<$}\;}

\slugcomment{8 September 2003; submitted to AJ}

\author{
Gordon T. Richards,\altaffilmark{1}
Michael A. Strauss,\altaffilmark{1}
Bartosz Pindor,\altaffilmark{1}
Zolt\'{a}n Haiman,\altaffilmark{1,2}
Xiaohui Fan,\altaffilmark{3}
Daniel Eisenstein,\altaffilmark{3}
Donald P. Schneider,\altaffilmark{4}
Neta A. Bahcall,\altaffilmark{1}
J. Brinkmann,\altaffilmark{5}
and Robert Brunner\altaffilmark{6}
}

\altaffiltext{1}{Princeton University Observatory, Peyton Hall, Princeton, NJ 08544.}
\altaffiltext{2}{Department of Astronomy, Columbia University, 550 West 120th Street, New York, NY 10027.}
\altaffiltext{3}{Steward Observatory, University of Arizona, 933 North Cherry Avenue, Tucson, AZ 85721.}
\altaffiltext{4}{Department of Astronomy and Astrophysics, The Pennsylvania State University, 525 Davey Laboratory, University Park, PA 16802.}
\altaffiltext{5}{Apache Point Observatory, P.O. Box 59, Sunspot, NM 88349.}
\altaffiltext{6}{Department of Astronomy, University of Illinois at Urbana-Champaign, 1002 West Green Street, Urbana, IL 61801-3080.}

\begin{document}

\title{A Snapshot Survey for Gravitational Lenses Among $z \ge 4.0$ Quasars: I. The $z>5.7$ Sample}

\begin{abstract}
Over the last few years, the Sloan Digital Sky Survey (SDSS) has
discovered several hundred quasars with redshift between 4.0 and 6.4.
Including the effects of magnification bias, one expects a priori that
an appreciable fraction of these objects are gravitationally lensed.
We have used the Advanced Camera for Surveys on the {\em Hubble Space
Telescope} to carry out a snapshot imaging survey of high-redshift
SDSS quasars to search for gravitationally split lenses.  This paper,
the first in a series reporting the results of the survey, describes
snapshot observations of four quasars at $z = 5.74, 5.82, 5.99$ and
6.30, respectively.  We find that none of these objects has a lensed
companion within 5 magnitudes with a separation larger than $0.3$
arcseconds; within 2.5 magnitudes, we can rule out companions within
0.1 arcseconds.  Based on the non-detection of strong lensing in these
four systems, we constrain the $z\sim6$ luminosity function to a slope
of $\beta>-4.63$ $(3\sigma)$, assuming a break in the quasar
luminosity function at $M_{1450}^{\star}=-24.0$.  We discuss the
implications of this constraint on the ionizing background due to
quasars in the early universe.  Given that these quasars are not
highly magnified, estimates of the masses of their central engines by
the Eddington argument must be taken seriously, possibly challenging
models of black hole formation.

\end{abstract}


\section{Introduction}

Before 1995, only 24 quasars with redshifts larger than four had been
published; most had been found in multicolor or grism surveys for
high-redshift quasars \markcite{whi+87,ssg87}(e.g., {Warren} {et~al.} 1987; {Schmidt}, {Schneider}, \& {Gunn} 1987).  In the past
eight years the number of known $z>4$ quasars has increased to more
than 400, largely due to the discoveries from the Sloan Digital Sky
Survey (SDSS; \markcite{yaa+00}{York} {et~al.} 2000) and the Digitized Palomar Sky Survey
\markcite{kdm+96}({Kennefick}, {Djorgovski}, \&  {Meylan} 1996).  The SDSS to date has published the discovery of over
200 quasars with redshifts greater than 3.6
\markcite{fsr+01,afr+01,sfs+01,sfh+03}({Fan} {et~al.} 2001b; {Anderson} {et~al.} 2001; {Schneider} {et~al.} 2001; {Schneider et al.} 2003).  In addition, \markcite{fwd+00}{Fan} {et~al.} (2000),
\markcite{fnl+01}{Fan} {et~al.} (2001a), and \markcite{fss+03}{Fan} {et~al.} (2003) have discovered the seven
highest-redshift quasars known, all with $z>5.7$ and four of which are
the subject of this paper.  This large sample provides us with new
opportunities to study high-redshift quasars; it has already been used
to determine the luminosity function of quasars and its evolution with
redshift from $z=3.6$ to $z=6.0$ \markcite{fss+01,fss+03}({Fan} {et~al.} 2001c, 2003).  The
luminosity function is well-fit with a power-law, while the comoving
number density of quasars drops exponentially with redshift over this
range.  One of the goals of this paper is to determine the extent to
which the observed luminosity function might be biased by
gravitational lensing.

Understanding the (intrinsic) quasar luminosity function and evolution
thereof at high redshifts is important for several reasons.  We
observe that supermassive, quiescent black holes are ubiquitous at low
redshift \markcite{mtr+98}({Magorrian} {et~al.} 1998); comparison of the present-day mass function
with the high-redshift luminosity function can constrain models for
the duty cycle of black holes and their feeding mechanisms.  There is
increasing evidence for a population of optically faint AGN at
high redshift, which appear in deep X-ray images with the {\em
Chandra} satellite \markcite{mcb+00,bha+01,abh+01}({Mushotzky} {et~al.} 2000; {Brandt} {et~al.} 2001; {Alexander} {et~al.} 2001); understanding their
nature requires that we understand the nature of the optically
luminous objects as well.  Finally, the optical (rest-frame UV)
luminosity function puts important constraints on the ultraviolet
ionizing background \markcite{fss+01}({Fan} {et~al.} 2001c), which is especially important now
that we appear to be observing the first hints that the universe
became completely reionized around redshift $z=6$
\markcite{bfw+01,dcs+01,wbf+03}(e.g., {Becker} {et~al.} 2001; {Djorgovski} {et~al.} 2001; {White} {et~al.} 2003).

The continuum luminosities of the $z>5.7$ quasars are all very high,
with absolute magnitudes at rest frame 1450$\,$\AA\ between $-26.8$
and $-28$ for the WMAP cosmology that we use throughout this paper (a
flat $\Lambda$ CDM model with $\Omega_m = 0.29$ and $H_0=72~{\rm
km~s^{-1}~Mpc^{-1}}$; \markcite{svp+03}{Spergel et al.} 2003).  This yields a lower limit to
the masses of their black holes of order $4 \times 10^9\,\rm M_\odot$
by the Eddington argument \markcite{fwd+00}({Fan} {et~al.} 2000).  It is a challenge to
explain in standard cosmologies how such massive black holes could
form at a cosmic epoch less than a billion years after the Big Bang
\markcite{er88,tur91,hl01}(e.g., {Efstathiou} \& {Rees} 1988; {Turner} 1991; {Haiman} \& {Loeb} 2001).

All these results assume that the apparent magnitudes that we measure
for these quasars are intrinsic to the quasar themselves, and
represent isotropic fluxes.  However, gravitational lensing can
systematically brighten the images of quasars by a significant amount.
This is a particularly important effect for flux-limited surveys like
SDSS, which are sensitive to the most luminous quasars at any given
redshift \markcite{mr93}(e.g., {Maoz} \& {Rix} 1993).  The resulting magnification bias
\markcite{pei95}(e.g., {Pei} 1995, and references therein) can systematically change
the quasar luminosity function, causing an overestimation of the black
hole masses powering these objects.

The expected fraction of multiply imaged quasars at a given redshift
depends both on the cosmological model (which sets the distribution
and profiles of halos as a function of redshift, as well as the
dependence of cosmological pathlength on redshift), and on the
luminosity function of quasars, which determines the effect of
magnification bias.  In the WMAP cosmology (using a halo distribution
taken from the large N-body simulations of \markcite{jfw+01}{Jenkins} {et~al.} 2001), the
fraction of random lines of sight at $z=4$ that produce multiple
images at all splitting angles (see the discussion of lensing models
in \S~3 below) is of order 0.2\%; this fraction rises to 0.4\% at
$z=6$.  However, magnification bias greatly increases the lensing rate
in a flux-limited sample.  At $z=4$, using the canonical cosmology,
and a luminosity function $\Phi(L) \propto L^{-2.5}$ \markcite{fsr+01}({Fan} {et~al.} 2001b)
with a break at $M_{1450}^\ast=-24.0$ (see \S~3 below for further
discussion), we expect about 1\% of quasars in our sample to be lensed
(specifically, multiply imaged with magnification of at least
two). This fraction increases to 8\% if we assume a steeper slope of
$\beta=3.5$, and to $\sim 2\%$ if we keep the slope of $\beta=2.5$,
but assume a break in $\Phi$ at a 10 times lower luminosity
($M_{1450}^\ast=-21.5$).  At $z=6$, magnification bias can result in
lensing probabilities approaching 100\% if the luminosity function is
particularly steep \markcite{chs+02,wl02,wl02b}({Comerford}, {Haiman}, \&  {Schaye} 2002; {Wyithe} \& {Loeb} 2002a, 2002b).  However, the detailed shape
of $\Phi$ is not well-measured, especially at higher redshifts and
luminosities.  Such quasars represent the highest peaks in the density
field at high redshift, where the mass function of massive dark halos
--- in which bright quasars likely reside --- is expected to be very
steep \markcite{fss+01}({Fan} {et~al.} 2001c).

The median expected splitting of gravitationally lensed quasars is
predicted to be somewhat less than $1\arcsec$ \markcite{tog84}({Turner}, {Ostriker}, \& {Gott} 1984), with flux
ratios as large as 20:1; this is in broad agreement with the observed
splitting separation found by the Cosmic Lens All Sky Survey (CLASS;
\markcite{bwj+03}{Browne} {et~al.} 2003), but see the discussion in \S 3 below on CLASS
implications for the details of the lens profiles.
The SDSS images themselves have PSF widths of order $1\arcsec -
1\farcs5$ \markcite{aaa+03}({Abazajian et al.} 2003), making them inadequate to look for all but
the very rare wide-separation ($\Delta \theta > 1\farcs5$) lenses
\markcite{ptl+03}({Pindor} {et~al.} 2003).  We are thus carrying out a snapshot imaging survey of
250 SDSS-discovered quasars with $z > 4.0$ using the Advanced Camera
for Surveys (ACS; \markcite{fch+03}{Ford} {et~al.} 2003) on the {\em Hubble Space Telescope
(HST)}, which can easily resolve pairs separated by as little as
$0\farcs1$ with flux ratios as large as 10:1.  Our observations should
therefore be sensitive to essentially all expected lenses
\markcite{tog84,hk87}({Turner} {et~al.} 1984; {Hinshaw} \& {Krauss} 1987).  \markcite{fss+03}{Fan} {et~al.} (2003) presented some preliminary results
based on some of these (and other, non-{\em HST}) data; here we give a
full analysis of the {\em HST} observations for all four of the
$z>5.7$ quasars in our sample.

Throughout this paper, we adopt the WMAP cosmology mentioned above,
with an rms mass fluctuation within a sphere of radius $8 \; h^{-1}$
Mpc of $\sigma_8=0.9$, and power--law index $n=0.99$ for the power
spectrum of density fluctuations \markcite{svp+03}({Spergel et al.} 2003). We also adopt the
cosmological transfer function from \markcite{eh99}{Eisenstein} \& {Hu} (1999).  Conversions between
$M_B$ and $M_{1450}$ assume $M_B = M_{1450} - 0.48$ \markcite{ssg95}({Schmidt}, {Schneider}, \& {Gunn} 1995) with
spectral index $\alpha_{\nu}=-0.5\;(f_{\nu}\propto\nu^{\alpha_{\nu}})$.

\section{The Data}

\subsection{Observations}

The full sample of $z > 4.0$ quasars included in our HST snapshot
program will be presented in future papers in this series.  In this
paper, we present results of imaging the four quasars with $z > 5.7$
discovered by \markcite{fwd+00,fnl+01}{Fan} {et~al.} (2000, 2001a); see Table~\ref{tab:tab1}.  These
four objects were selected from the SDSS imaging database based on
their extremely red $i-z$ color, absence of detection in $u,g,$ and
$r$, and relatively blue $z-J$ color (which distinguishes the $z >
5.7$ quasars from brown dwarfs).

The SDSS images of these quasars have an image scale of
$0\farcs396\,{\rm pixel}^{-1}$ and $z$-band seeing FWHM between
$1\farcs4$ and $1\farcs8$.  In order to test the ability of the SDSS
photometric pipeline to identify quasar pairs, we created simulated
SDSS images of pairs of point sources (see \markcite{ptl+03}{Pindor} {et~al.} 2003 for
details) in these observing conditions and at the appropriate
signal-to-noise ratio.  We find that the SDSS star-galaxy separator
would have identified pairs of point sources as resolved objects for
image separations greater than $\sim 1 \arcsec$ and for flux ratios
less than $\sim$ 5:1.  Thus the SDSS data are suitable for exploring
only a fraction of the parameter space of lensing (in terms of
separations and flux ratios) that is of interest; higher resolution
{\em HST} observations were needed to probe the rest of that parameter
space.  We did not restrict our sample to point sources since the SDSS
star-galaxy separation algorithm (see \markcite{sjd+02}{Scranton} {et~al.} 2002) breaks down at
these faint magnitudes.  The only object of the four that was flagged
as an extended source was SDSSp~J104433.04-012502.2.

We used the High Resolution Camera (HRC) on ACS to image the four
$z>5.7$ quasars presented herein.  For these four objects, the
Ly$\alpha$ forest lies in the SDSS $i$-band; we therefore observed
them in the SDSS $z$-band, F850LP, despite its lower sensitivity.  The
Wide-Field Camera on ACS has higher sensitivity in $i$ and $z$ than
does the HRC, but has has substantially higher overhead and is mildly
undersampled.  The exposure times were 1200 seconds for each object,
600 seconds in each of two exposures to help in cosmic ray rejection.
We did not dither the images, and our cosmic ray rejection is far from
perfect (\S~2.2); however, our analysis indicates that imperfect
cosmic ray removal is not critical for our search for gravitational
lenses.

\subsection{Data Processing}

\begin{figure}[p]
\epsscale{1.0}
\plotone{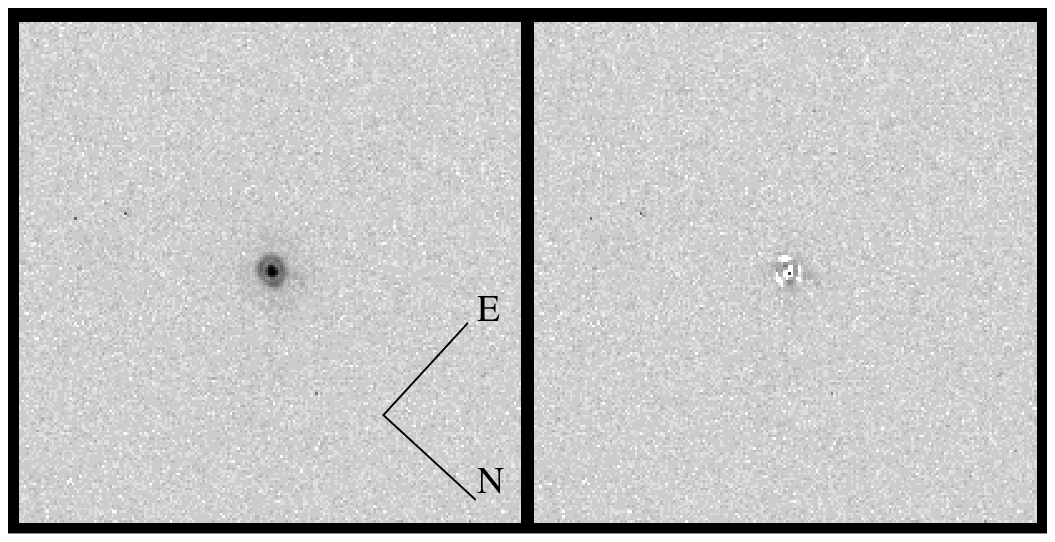}
\caption{{\em HST}/ACS/HRC F850LP images of SDSS J1044-0125
($z=5.74$).  The scale is $5\arcsec\times5\arcsec$ in each of the
panels. The left hand panel is the ``cosmic ray rejected'' (CRJ)
output of CALACS.  The right hand panel is the same CRJ image after
subtraction of the Tiny Tim PSF (v6.0).  Note that the $m_B \approx
25$ faint galaxy that \markcite{stm+03}{Shioya} {et~al.} (2002) detected $1\farcs9$ to the
south-west of the quasar position is not visible at the depth of this
image.
\label{fig:fig1}}
\end{figure}

The raw images were calibrated by the CALACS package in
IRAF\footnote{IRAF is distributed by the National Optical Astronomy
Observatories, which are operated by the Association of Universities
for Research in Astronomy, Inc., under cooperative agreement with the
National Science Foundation.} as part of on-the-fly-reprocessing
(OTFR) at the time of download.  For the most part, we find these
initial reductions to be sufficient for our purposes.  The images that
we present are the ``cosmic ray rejected'' (CRJ) images that are
output by the OTFR algorithms at STScI.  These files were re-requested
a number of weeks after the observations were completed to ensure that
we are using the latest ``reference'' files and that the OTFR
reprocessing is current.  The CRJ files have all been reduced in the
standard manner, including having been overscan-, bias- and
dark-corrected, flat-fielded and photometrically calibrated, in
addition to having bad pixels masked and cosmic rays removed (see the
ACS
manual\footnote{http://www.stsci.edu/hst/acs/documents/handbooks/cycle12/cover.html}
for more details).  The plate scale on the HRC of the ACS is not the
same in the $x$ and $y$ directions and the sky axes are not
perpendicular on the CCD; these CRJ images have {\em not} been
corrected for this distortion.  A $5\arcsec\times5\arcsec$ cutout of
each of the ACS images is given in
Figures~\ref{fig:fig1}-\ref{fig:fig4}.

With the exception of cosmic ray rejection and hot pixel
identification, the images output by the OTFR algorithm are acceptable
for our purposes.  The CALACS rejects pixels that are significantly
different in the two images.  This process successfully rejects the
majority of the cosmic rays in our images.  However, many of the
cosmic rays occupy multiple pixels, and we have only two exposures,
meaning that some cosmic rays remain.  In addition, hot pixels can
remain in the corrected images if they have not been properly
identified in the masks that are constructed for each period of
observations between the annealing procedure that helps to repair hot
pixels.  Fortunately, the morphologies of the residual cosmic rays and
hot pixels are considerably different from that of the ACS PSF (the
PSF is spread out over many tens of pixels), so spurious sources due
to cosmic rays and most other defects are easy to recognize.

\subsection{Subtracting the Point Source}

The point spread function is narrow enough that any lens with a
separation greater than $\sim0\farcs2$ will be obvious by visual
inspection (see \S~\ref{sec:multiple}).  To search for faint secondary
images at separations smaller than this requires that we fit and
subtract off a model for the point spread function of each image.  We
have done this using the v6.0 (pre-launch of ACS) version of the Tiny
Tim software
\markcite{kri95}({Krist} 1995)\footnote{http://www.stsci.edu/software/tinytim/}, which
produces a model PSF for the instruments on {\em HST} given the
object's SED and position in the focal plane, the filter curve, and
knowledge of the optics of the instrument.  Such model PSFs are
necessary because it is not always possible to do PSF subtraction using
an empirical template (another point source in the field-of-view).
That is, given the positional dependence of the PSF across the
field-of-view, the Tiny Tim models are often better than using an
empirical template.  For our ACS/HRC images, the model correctly takes
into account the different scales and non-orthogonality of the $x$ and
$y$ axes.

We fit the PSF model to each CRJ image, allowing the location on the
CCD and normalization to vary, and using sinc interpolation when
shifting the model PSF by fractional pixels.  Inputs to Tiny Tim were
the camera used (ACS/HRC), the location on the CCD of each quasar to
the nearest pixel (as measured by {\tt imexam} in IRAF) and a Keck
spectrum \markcite{bfw+01}({Becker} {et~al.} 2001) of each quasar (for the wavelength dependence
of the PSF).  We minimized the sum of the square of residuals (since
the formal errors per pixel over the image are close to constant).
The results are shown in Figures~\ref{fig:fig1}--\ref{fig:fig4}.  Each
of the four images show the familiar first Airy ring; on a very hard
stretch, the second Airy ring is faintly visible.  Note that {\em HST}
PSFs are not symmetric as a result of the three primary mirror
supports, the signature of which can be seen in the PSF subtracted
images.  Given inevitably somewhat imperfect PSF subtraction and the
expected asymmetry of the residual images, in none of the four cases
is there an obvious secondary image; none of them are obviously
gravitationally lensed.

\subsection{Notes on Individual Objects}

\subsubsection{SDSSp J104433.04-012502.2; $z=5.74$}

The ACS/HRC image of this quasar is shown in the left panel of
Figure~\ref{fig:fig1}; the right panel shows the residuals after PSF
subtraction.  The SDSS $z$-band magnitude of this quasar is 19.3; the
{\em HST} $z$-band (ST) magnitude is 19.91, see also
Table~\ref{tab:tab1}.  The offset between the SDSS and ST $z$-band
magnitude is systematic (see the other objects below) with the ST
magnitudes always being fainter.  This is probably due to the
differences in the filter curves (the SDSS curve having less red
response due to atmospheric features) and the differences between a
Vega-based (ST) and an AB-based (SDSS) photometric system given that
Vega has an AB $z$-band magnitude of 0.57 \markcite{fig+96}({Fukugita} {et~al.} 1996).  We find no
offset between the SDSS and ST $i$-band magnitudes (in the {\em HST}
images of our lower redshift quasars), where the filter curves appear
to be much more similar.

\markcite{fwd+00}{Fan} {et~al.} (2000) describe a $K'$ image of this object taken with the
Keck 10m telescope in $0\farcs375$ seeing, which was consistent with a
point source.  That is consistent with the HST image; there is no
apparent splitting in this object.

\markcite{stm+03}{Shioya} {et~al.} (2002) found a faint galaxy ($m_B \approx 25$) $1\farcs9$
south-west of this quasar.  Our 20-minute exposure shows this object
faintly after rebinning.  \markcite{stm+03}{Shioya} {et~al.} (2002) suggest that this galaxy
could amplify the flux of the quasar by up to a factor of two.  The
lack of a secondary image of the quasar restricts any possible
magnification to be no larger than this --- in the absence of
microlensing and external shear \markcite{wl02}({Wyithe} \& {Loeb} 2002a).

\subsubsection{SDSSp J083643.85+005453.3; $z=5.82$}

\begin{figure}[p]
\epsscale{1.0}
\plotone{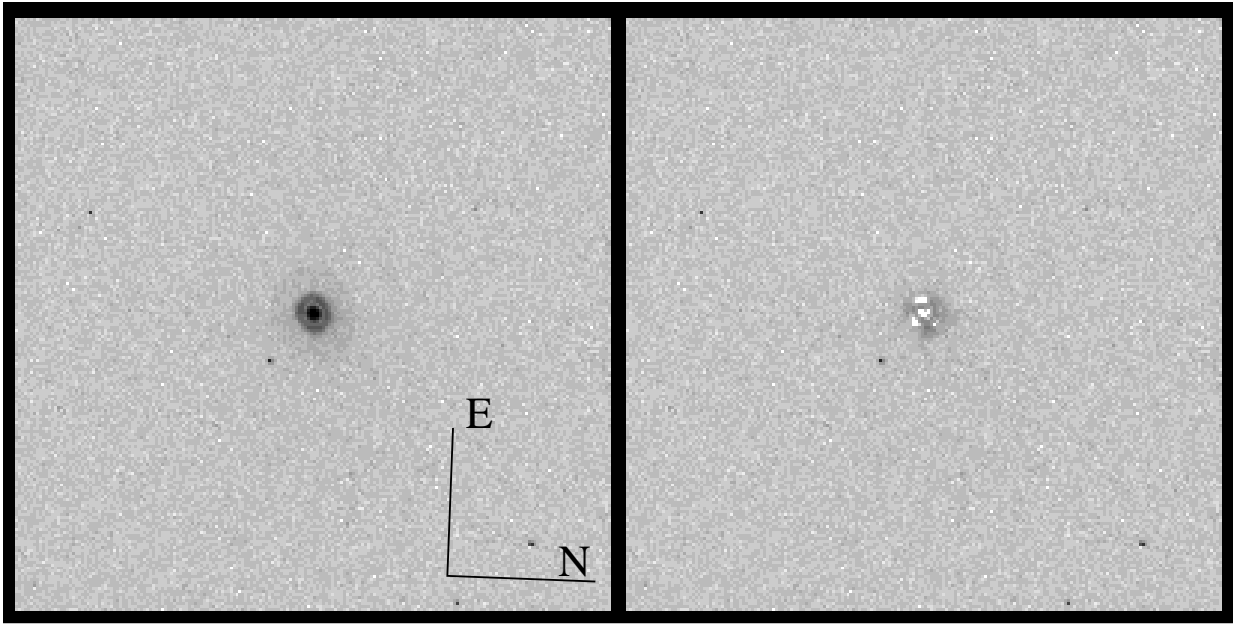}
\caption{As in Figure 1, for SDSS J0836+0054 ($z=5.82$). 
\label{fig:fig2}}
\end{figure}

The ACS/HRC image of this quasar and residuals after PSF subtraction
are shown in Figure~\ref{fig:fig2}.  The SDSS $z$-band magnitude of
this quasar is 18.79; the {\em HST} $z$-band (ST) magnitude is 19.49,
see also Table~\ref{tab:tab1}.  One other point source and two other
extended sources are in the ACS/HRC field of this quasar, but all are
more than $5\arcsec$ away (out of the field-of-view of
Fig.~\ref{fig:fig2}), and are therefore unlikely to lens (or
significantly magnify) the quasar.  The black dots in the
field-of-view of Figure~\ref{fig:fig2} are residual cosmic rays that
were not properly cleaned by the OTFR reductions.

\subsubsection{SDSSp J130608.26+035626.3; $z=5.99$}

\begin{figure}[p]
\epsscale{1.0}
\plotone{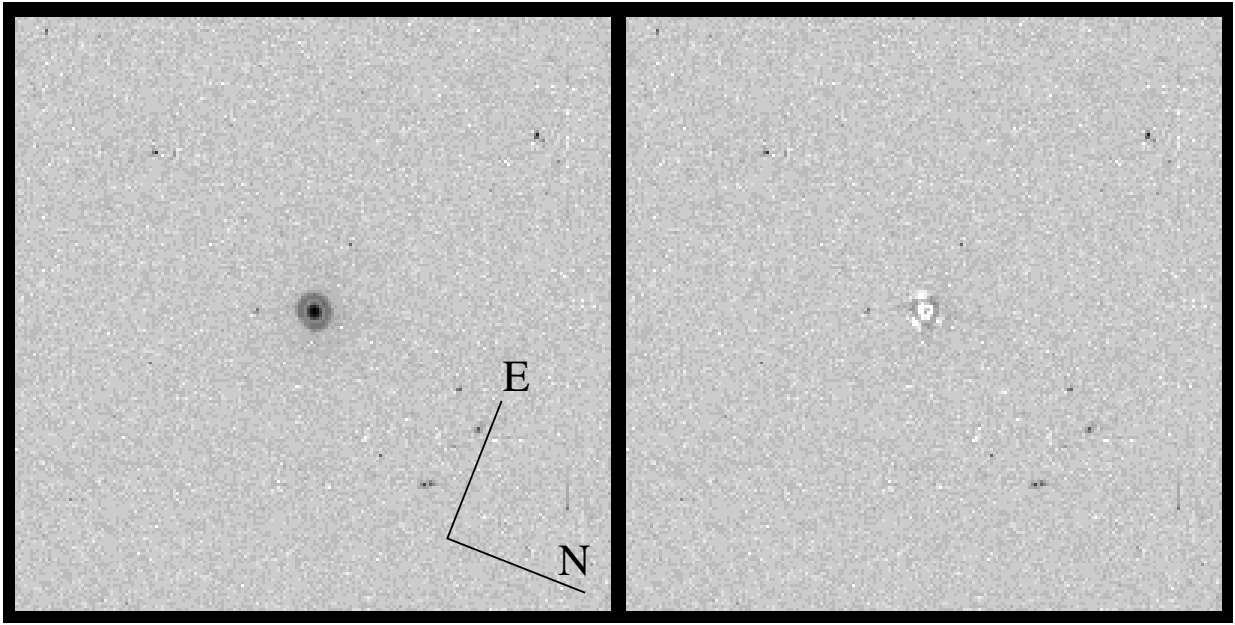}
\caption{As in Figure 1, for SDSS J 1306+0356 ($z=5.99$). 
\label{fig:fig3}}
\end{figure}

The ACS/HRC image of this quasar is shown in Figure~\ref{fig:fig3}.
The SDSS $z$-band magnitude of this quasar is 19.46; the {\em HST}
$z$-band (ST) magnitude is 20.23, see also Table~\ref{tab:tab1}.
There is some evidence for a few very faint extended sources in the
ACS/HRC field (off the image in Fig.~\ref{fig:fig3}), but as with
Figure~\ref{fig:fig2}, all of the black dots in Figure~\ref{fig:fig3}
are residual cosmic rays.  \markcite{yst+03}{Yamada et al.} (2003) have taken deep images to $i
\approx 25.5$ in moderate seeing with Subaru in this field, and have
found no foreground galaxies close enough to the line of sight to
cause gravitational lensing.

\subsubsection{SDSSp J103027.10+052455.0; $z=6.30$}

\begin{figure}[p]
\epsscale{1.0}
\plotone{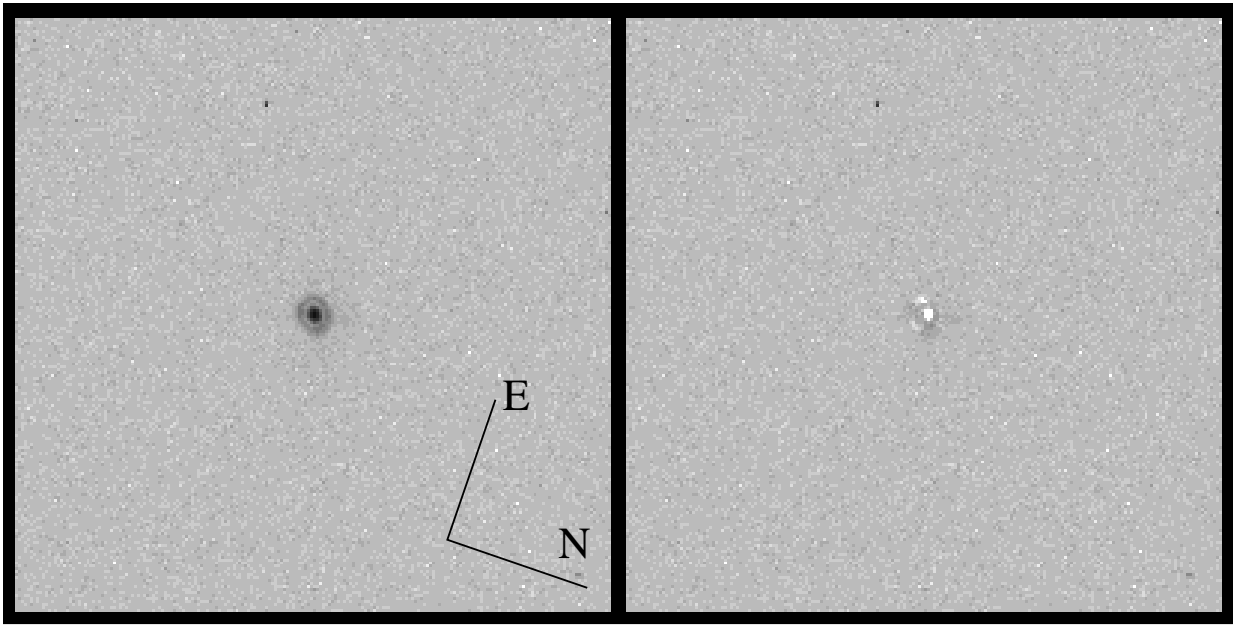}
\caption{As in Figure 1, for SDSS J1030+0524 ($z=6.30$).  
\label{fig:fig4}}
\end{figure}

The ACS/HRC image of this quasar is shown in Figure~\ref{fig:fig4}.
The SDSS $z$-band magnitude of this quasar is 20.02; the {\em HST}
$z$-band (ST) magnitude is 20.54, see also Table~\ref{tab:tab1}.
There is one other relatively bright point source in the field, nearly
due west of the target quasar, but it is at the edge of the field (off
the image in Fig.~\ref{fig:fig4}) and is not of interest in terms of
lensing.  As in the previous object, a deep image by \markcite{yst+03}{Yamada et al.} (2003)
shows no obvious lensing galaxies in the foreground.  However,
\markcite{pcb+03}{Petric} {et~al.} (2003) have taken deep VLA images at 20 cm in this field, and
found a statistically significant overdensity of four radio sources
within $1'$ of the quasar, which may indicate the possibility of a
magnifying cluster along the line of sight.

\subsection{Looking for Multiple Images}
\label{sec:multiple}

Our ACS images have an image scale of $0\farcs025\,{\rm pixel}^{-1}$,
and we can visually rule out any multiple imaging at scales larger
than about $0\farcs2$.  To make this more quantitative, we have
carried out simulations of quasar pairs at various separations and
flux ratios, by adding a scaled-down and shifted version of our quasar
images to itself.  This naive procedure produces a slightly incorrect
sky level, but the inaccuracy is negligible for our present purposes.
Figure~\ref{fig:sim3} shows simulated observations of SDSS~J0836+0054
with an additional point source 2.5 and 5.0 mag fainter than the
quasar and separated from the quasar by $0\farcs3$.  The secondary
image is obvious in both cases even without PSF subtraction.
Figure~\ref{fig:sim1} shows the result of a $0\farcs1$ offset.  With a
flux ratio of 10:1, the secondary object is visible as an enhancement
in the first Airy ring of the primary object; it becomes clear upon
PSF subtraction (Figure~\ref{fig:sim1resid}).  With a flux ratio of
100:1, we cannot discern a pair with $0\farcs1$ separation even after
subtraction of the PSF; however, even if we were able to do so, it
would not significantly improve the constraints that we derive below
since most split images will have larger image separations.

\section{Constraining the Slope of the Quasar Luminosity Function}

The shape of the high-redshift quasar luminosity function is not
tightly constrained because of the limited size and luminosity range
of the available statistical samples.  \markcite{ssg95}{Schmidt} {et~al.} (1995), using a set of
90 quasars with $2.7 < z < 4.8$, found a power-law luminosity function
slope ($\beta$) of $\approx-2$.  Fan et al. (2001c) measured $\beta
\approx -2.5\pm0.3$ from 39 quasars in the redshift interval 3.6 to
5.0.  These slopes apply to the high-luminosity ($M \lesssim -26$)
part of the luminosity function.  At these high redshifts, direct
observations currently yield only poor constraints on the slope of the
quasar luminosity function (LF) at $z\sim 5$, and essentially no
constraints at $z\gsim 6$, where the luminosities of the known quasars
all lie within $\sim0.5$ magnitudes of the detection threshold of
SDSS. From the number counts of $z\sim6$ quasars, \markcite{fss+01}{Fan} {et~al.} (2001c) place
a 3-$\sigma$ lower limit on the slope of $\beta=-4.6$.  Gravitational
lensing tends to flatten the intrinsic luminosity function, producing
an apparent slope of $\beta=-3$ in the limit of an arbitrarily steep
intrinsic LF down to a fixed low--luminosity cutoff
\markcite{sef92}(e.g., {Schneider}, {Ehlers}, \& {Falco} 1992).  As a result, the relatively shallow apparent
slope at $z\sim 5$ does not necessarily reflect the slope of the
underlying LF, which could be much steeper, if most quasars were
lensed and magnified.

One of the primary goals of this work is to constrain the slope of the
quasar LF at high redshift, using the lack of evidence for strong
lensing among the four quasars in the HST images \markcite{chs+02}({Comerford} {et~al.} 2002).  In
line with previous work, we describe the intrinsic (not necessarily
observed) quasar luminosity function as a broken power law
\markcite{bsp88,pei95}(e.g., {Boyle}, {Shanks}, \& {Peterson} 1988; {Pei} 1995):
\begin{equation}
\Phi_{\rm int}(L)=\frac{\Phi_\ast/L_\ast}{
(L/L_\ast)^{\beta_l}+(L/L_\ast)^{\beta_h}}.
\label{eq:qlf}
\end{equation}
The LF is described by four parameters: the normalization $\Phi_\ast$,
the faint-end slope $\beta_l$, the bright-end slope $\beta_h$, and the
characteristic luminosity $L_\ast$ at which the LF steepens. The
lensing probability is very sensitive to the last two parameters,
$\beta_h$ and $L_\ast$, while it is only weakly dependent on the
faint--end slope below the break, and it is strictly independent of
the normalization $\Phi_\ast$.  Here we apply the lensing model from
\markcite{chs+02}{Comerford} {et~al.} (2002), in which lenses are associated with dark matter halos,
to compute the total lensing probability, including the effect of
magnification bias.  In this model, the abundance of halos as a
function of potential well depth is adopted from the simulations of
\markcite{jfw+01}{Jenkins} {et~al.} (2001). While it would be simplest to assume that the halos
have a common density profile, such as the singular isothermal sphere
(SIS), recent work has suggested that the typical density profile of
massive lenses is shallower than that of the SIS
\markcite{pm00,kw01,lo02,hm03}({Porciani} \& {Madau} 2000; {Kochanek} \& {White} 2001; {Li} \& {Ostriker} 2002; {Huterer} \& {Ma} 2003).  The observed dearth of large--separation
lenses cannot be explained if all lenses have the same density
profile.  A prescription that fits the observed steep splitting angle
distribution is to assume that all halos below $M\approx 10^{13}~{\rm
M_\odot}$ have SIS profiles (adopting a standard conversion between
circular velocity and halo mass), while all halos above this mass
follow the dark matter density profile suggested by \markcite{nfw97}{Navarro}, {Frenk}, \& {White} (1997, hereafter
NFW).  NFW profiles are much less efficient lenses than are SIS
profiles.  This prescription is physically motivated in \markcite{kw01}{Kochanek} \& {White} (2001)
based on the efficiency of gas cooling, and we adopt it in this
paper. In our calculations, this is essentially equivalent to ignoring
all lenses above a halo mass of $10^{13} {\rm M_\odot}$, as the
massive halos do not contribute to lensing at small separations.

As an example, we first fix the characteristic luminosity (``knee'')
of the quasar LF at a fiducial value of $M_{1450}^\ast=-24.0$, and
derive constraints on the bright-end slope, $\beta_h$.  The choice for
this value is somewhat arbitrary: the highest redshift where the
quasar LF is measured to faint enough magnitudes to define a reliable
break is at $z\approx 2.15$ \markcite{bsc+00}({Boyle} {et~al.} 2000), where
$M_{1450}^\ast\approx -25.6\;(M_{B}^\ast \approx - 26.1)$. In pure
luminosity evolution models, the break at $z\approx 5$ would then be
at around $M_{1450}^\ast=-24.0$ \markcite{mhr99,wl02}(e.g. {Madau}, {Haardt}, \& {Rees} 1999; {Wyithe} \& {Loeb} 2002a).  Having
fixed the value of the break, we next vary the slope $\beta_h$ of the
LF, and require that the expected probability for no lensing among the
four highest redshift quasars is P=0.0026 ($=3\sigma$) or greater.
Assuming that we can rule out a second image with a flux ratio of 100
down to a splitting angle of $0\farcs3$, we find the constraint on the
slope $\beta_h>-4.63$.  We find that resolving high--contrast lensing
configurations, with a flux ratio $>10$, or resolving angular
separations smaller than $0\farcs3$ does not significantly improve
this constraint, since in any model, the fraction of lenses with
separations less than $0\farcs3$ is small.  For example, ruling out
flux ratios of 10 down to $0\farcs1$ would only improve the constraint
to $\beta_h>-4.58$.

It is more informative to illustrate joint constraints on the
parameters $(M_{1450}^\ast, \beta_h)$.  In Figure~\ref{fig:contours},
we show contours for the likelihood at $P=0.32$, 0.05, 0.01, and 0.001
that none of the four $z>5.7$ quasars are lensed (again, to a
separation of 0.3 arcseconds, and to flux ratios of 100).  The figure
shows, as in \markcite{fss+03}{Fan} {et~al.} (2003), that the derived limit on the slope
$\beta_h$ depends strongly on the assumed break.  Overall, the
constraint shown in this figure is similar to that derived from the
sample of seven high--redshift quasars in \markcite{fss+03}{Fan} {et~al.} (2003), although the
present constraints are somewhat weaker, due to the smaller number of
sources.  The recent results of \markcite{wyi03}{Wyithe} (2003) are consistent with ours
under the assumption of no lensing.  As shown in \markcite{wyi03}{Wyithe} (2003), a
stronger constraint is available if two of the known $z>6$ quasars are
indeed magnified, as suggested by \markcite{stm+03}{Shioya} {et~al.} (2002) and \markcite{wbf+03}{White} {et~al.} (2003).

\section{Discussion}

The lack of lenses among the four high--redshift quasars allows us to
place constraints on the quasar LF at high redshift, shown in
Figure~\ref{fig:contours}.  It is useful to keep in mind that only
models with extreme magnification bias, predicting significant lensing
probabilities, can be constrained by the {\it lack} of lensing events.
For example, for the range of LF slopes considered by \markcite{wl02}{Wyithe} \& {Loeb} (2002a),
$\beta_h=2.58-3.43$, the lensing fraction for the $z\sim6$ SDSS sample
would be 7-30\%.  The chance of finding no lenses among the four
quasars for their steepest slope of 3.43 is $\sim (1-0.3)^4=0.24$,
which translates to a significance of only $\sim1\sigma$ for ruling
out this slope (consistent with our results shown in
Figure~\ref{fig:contours} at the break of $M_{1450}=-24.0$).

Accordingly, as Figure~\ref{fig:contours} shows, the constraints
obtained here still allow very steep LF slopes, especially if we
assume that the break in the LF lies at relatively bright magnitudes.
Nevertheless, these constraints are fairly direct, and are among the
most stringent limits we currently have on the shape of the
high-redshift quasar LF.  We expect that the full $z>4$ sample will
further tighten the constraints (or yield a measurement should a lens
be found).  Table~\ref{tab:tab2} shows that the lensing probability
(for a slope of $\beta=-3.5$ and break $M_{1450}=-24.0$) decreases
from 8.2\% at $z=6$ to 1.2\% at $z=4$.  This implies that seeing no
lensing with a single $z=6$ quasar provides a constraint that is
equivalent to about seven $z=4$ quasars, from the equation
$(1-0.012)^N = (1-0.082)$. The lensing probability is a steep function
of $\beta_h$ and $M^\ast$, but since our full sample will include
$\sim250/4 \approx 60$ times more sources than utilized here, we
expect that we will be able to significantly tighten the constraints
in the $(\beta_h,M^\ast)$ plane with this sample.

Given our assumed shape of the quasar LF, the constraints obtained in
the $(\beta_h,M^\ast)$ plane translate directly to the global ionizing
emissivity of the $z\sim 6$ quasar population, $\nu \epsilon_\nu=\int
L \Phi(L) dL$.  The emissivity is of great interest, since a related
quantity --- the mean background flux --- can be constrained
independently from the level of resonant neutral hydrogen absorption
detected in high--resolution spectra of the SDSS quasars.
Specifically, recent upper limits on this background were obtained
from the Gunn--Peterson (1965) trough of the $z=6.30$ quasar
\markcite{mm01,cm02,lhz+02,fns+02}(e.g., {McDonald} \& {Miralda-Escud{\' e}} 2001; {Cen} \& {McDonald} 2002; {Lidz} {et~al.} 2002; {Fan} {et~al.} 2002), requiring that
$\Gamma<10^{-13}$ ionizations per hydrogen atom per second at $z=6$.
The conversion to the ionizing emissivity $\nu \epsilon_\nu$ is
\begin{equation}
\Gamma=\frac{\pi\sigma_H}{h_{\rm P}}\lambda_{\rm mfp}\frac{(1+z)^3}{4\pi}
\frac{\nu\epsilon_\nu}{\nu_{\rm H}}~~~~{\rm s^{-1}}, 
\end{equation}
where we have assumed $\epsilon_\nu\propto\nu^{-1}$,
$\sigma_H=6.3\times 10^{-18}~{\rm cm^{-2}}$ is the hydrogen ionization
cross section at the Lyman limit (and we assumed $\sigma_H(\nu)\propto
\nu^{-3}$ above the threshold), $h_{\rm P}$ is the Planck constant,
$\nu_{\rm H}=3.29\times 10^{15}$Hz is the frequency of a Lyman limit
photon, and $\lambda_{\rm mfp}$ is the mean free path of ionizing
photons (see the Appendix in \markcite{chs+02}{Comerford} {et~al.} 2002).  The last quantity
unfortunately has a very large uncertainty.  Nevertheless, for a crude
estimate, we here assume that the effective mean free path corresponds
to a redshift interval of $\Delta z=0.17$ at $z=3$
\markcite{hm96,spa01}(see {Haardt} \& {Madau} 1996; {Steidel}, {Pettini}, \&  {Adelberger} 2001), and scales with redshift as $\lambda_{\rm
mfp}\propto (1+z)^{-6}$ towards higher redshift \markcite{cm02}({Cen} \& {McDonald} 2002).  In this
case, the model with $M_{1450}^\ast=-24.0$ and $\beta_h=-4.6$
corresponds to $\Gamma\sim 10^{-13}\,\rm sec^{-1}$.  In other words,
the constraint we derive here from the absence of lensing is
approximately at the same level as that implied by the upper limit on
the ionizing background.  Note that the constraints we obtain apply to
the total emissivity of quasars alone, but the upper limits on the
background from the Gunn-Peterson trough constrains the total
emissivity (including galaxies).  Future improvements of the lensing
constraint can yield tighter limits on the ionizing emissivity of the
quasar population than hitherto possible --- subject to the
uncertainties in (1) the mean evolution of the mean free path with
redshift, (2) the faint--end slope $\beta_l$, which does affect the
total emissivity $\nu\epsilon_\nu$, and (3) the escape fraction of
ionizing photons at $z\sim 6$.

The fact that none of the four high--redshift quasars appears to be
strongly magnified also implies that the masses inferred for the
central black holes from the Eddington limit argument are real.  If a
quasar is strongly beamed towards us, the bolometric luminosity would
be overestimated, but \markcite{wmj03}{Willott}, {McLure}, \& {Jarvis} (2003), following \markcite{hc02}{Haiman} \& {Cen} (2002) argued
that emission line equivalent widths being normal mitigates the
beaming hypothesis.  The inferred black hole masses in these four
quasars are very high, $ 2-4 \times 10^9~{\rm M_\odot}$; how black
holes can grow so large so early in the Universe requires explanation
\markcite{er88,tur91}({Efstathiou} \& {Rees} 1988; {Turner} 1991).  \markcite{hl01}{Haiman} \& {Loeb} (2001) considered the presence of the $\sim
4\times10^9~{\rm M_\odot}$ black hole for the $z=5.75$ quasar in
simple models for black hole growth. They found that it could be
marginally accommodated in simple models of black hole growth by
mergers and accretion in an LCDM universe. These considerations depend
strongly on the redshift, and on the typical radiative efficiency of
black holes. The highest redshift quasars therefore represent a bigger
challenge to models, especially if their radiative efficiency during
their growth was as high as 20\%, the typical value inferred recently
for the brightest quasars at lower redshifts \markcite{yt02}({Yu} \& {Tremaine} 2002).  This is
twice the fiducial value of $\sim10\%$ adopted in \markcite{hl01}{Haiman} \& {Loeb} (2001) and
most other previous work on black hole growth, doubling the expected
growth timescale.

Finally, the lack of strong lensing appears consistent with the large
apparent sizes of ionized regions around the $z\sim 6$ quasars.
\markcite{hc02}{Haiman} \& {Cen} (2002) argue that the $z=6.30$ source cannot be magnified by
more than a factor of a few, since an intrinsically faint source would
not be able to ionize a neutral IGM out to the observed distance away
from the source of $\sim 30$ (comoving) Mpc --- the spectrum shows
transmission at wavelengths corresponding to this distance.  Similar
conclusions were drawn for the $z=6.42$ quasar \markcite{fss+03}({Fan} {et~al.} 2003) recently
by \markcite{wbf+03}{White} {et~al.} (2003).

\section{Conclusions and Future Work}

We have obtained high resolution {\em HST} images of the four highest
redshift quasars known prior to 2003 to look for the signature of
gravitational lensing.  We have found no evidence of multiple images,
significantly limiting the amount by which these quasars can be
magnified by foreground mass concentrations (in the absence of
microlensing).  Thus the masses of these quasars as derived from the
Eddington argument cannot be significantly inflated by magnification.

The lack of any strong lenses puts a $3\sigma$ constraint on the
bright end slope of the $z>5$ luminosity function of $\beta_h>-4.63$.
This limit on the slope of the luminosity function is also consistent
with previous limits on the ionizing emissivity of $z\sim6$ quasars of
$\Gamma\sim 10^{-13}\,\rm sec^{-1}$ ionizations per hydrogen atom per
second based on the Gunn-Peterson trough in the highest redshift
quasar studied herein.

The next paper in the series will present an analysis of a much larger
sample of $z>4$ quasar observations with {\em HST} and will discuss
the constraints that we can place upon the quasar luminosity function
from that larger (but lower redshift) sample.  Furthermore, there are
many more $z>4$ (and even a few $z>5.7$) quasars known now than were
known at the time that these {\em HST} observations were approved.
Higher resolution images should be taken of as many of these as
possible; much of the parameter space can even be covered by
ground-based observations since {\em HST} resolution splittings are
only needed to discern the smallest expected splittings.

\acknowledgements

Funding for the creation and distribution of the SDSS Archive has been
provided by the Alfred P. Sloan Foundation, the Participating
Institutions, the National Aeronautics and Space Administration, the
National Science Foundation, the U.S. Department of Energy, the
Japanese Monbukagakusho, and the Max Planck Society. The SDSS Web site
is http://www.sdss.org/.  The SDSS is managed by the Astrophysical
Research Consortium (ARC) for the Participating Institutions. The
Participating Institutions are The University of Chicago, Fermilab,
the Institute for Advanced Study, the Japan Participation Group, The
Johns Hopkins University, Los Alamos National Laboratory, the
Max-Planck-Institute for Astronomy (MPIA), the Max-Planck-Institute
for Astrophysics (MPA), New Mexico State University, University of
Pittsburgh, Princeton University, the United States Naval Observatory,
and the University of Washington.  This work was supported in part by
HST grant HST-GO-09472.01-A (G.~T.~R. and M.~A.~S.), National Science
Foundation grants AST-9900703 (D.~P.~S.), AST-0071091 (M.~A.~S.),
AST-0307384 (X.~F.), AST-0307291 (Z.~H.) and AST-0307200 (Z.~H.).
X.~F. and D.~J.~E. acknowledge Alfred P. Sloan Research Fellowships.

\clearpage

\nocite{gp65}



\begin{figure}[p]
\epsscale{1.0}
\plotone{RichardsGT_hstlens.fig5.ps}
\caption{Simulated lenses using the SDSS J0836+0054 images.  ({\em
Left}) Lens simulated by scaling down the PSF of SDSS J0836+0054 by a
factor of 10 and shifting it down by $0\farcs3$.  ({\em Right}) Lens
simulated by scaling down the PSF of SDSS J0836+0054 by a factor of
100 and shifting it down by $0\farcs3$.\label{fig:sim3}}
\end{figure}

\begin{figure}[p]
\epsscale{1.0}
\plotone{RichardsGT_hstlens.fig6.ps}
\caption{Simulated lenses using the SDSS J0836+0054 images.  ({\em
Left}) Lens simulated by scaling down the PSF of SDSS J0836+0054 by a
factor of 10 and shifting it down by $0\farcs1$.  ({\em Right}) Lens
simulated by scaling down the PSF of SDSS J0836+0054 by a factor of
100 and shifting it down by $0\farcs1$.\label{fig:sim1}}
\end{figure}

\begin{figure}[p]
\epsscale{1.0}
\plotone{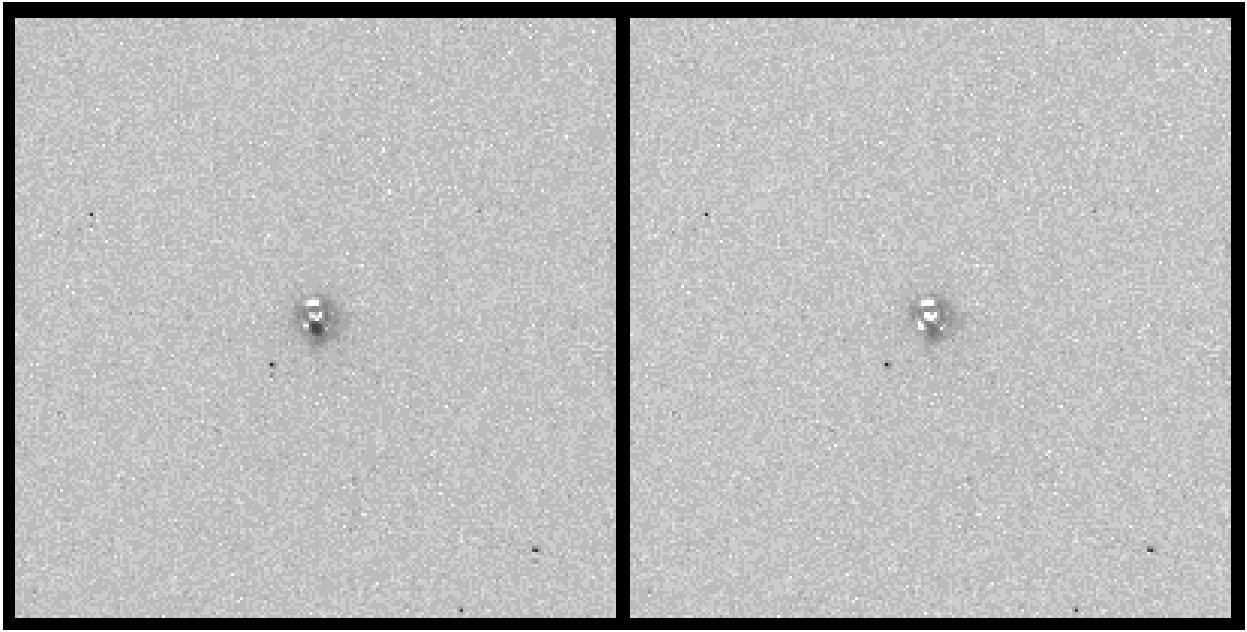}
\caption{The same as Figure~\ref{fig:sim1}, but with PSF subtraction
of the brightest component.  ({\em Left}) Lens simulated by scaling
down the PSF of SDSS J0836+0054 by a factor of 10 and shifting it down
by $0\farcs1$.  ({\em Right}) Lens simulated by scaling down the PSF
of SDSS J0836+0054 by a factor of 100 and shifting it down by
$0\farcs1$.  For the case on the left, we can clearly see the second
source, despite the PSF residuals (note the triangular pattern due to
the three primary mirror supports).  The second image on the right is
not possible to resolve from the primary image without better PSF
subtraction (if at all).\label{fig:sim1resid}}
\end{figure}

\begin{figure}[p]
\epsscale{1.0}
\plotone{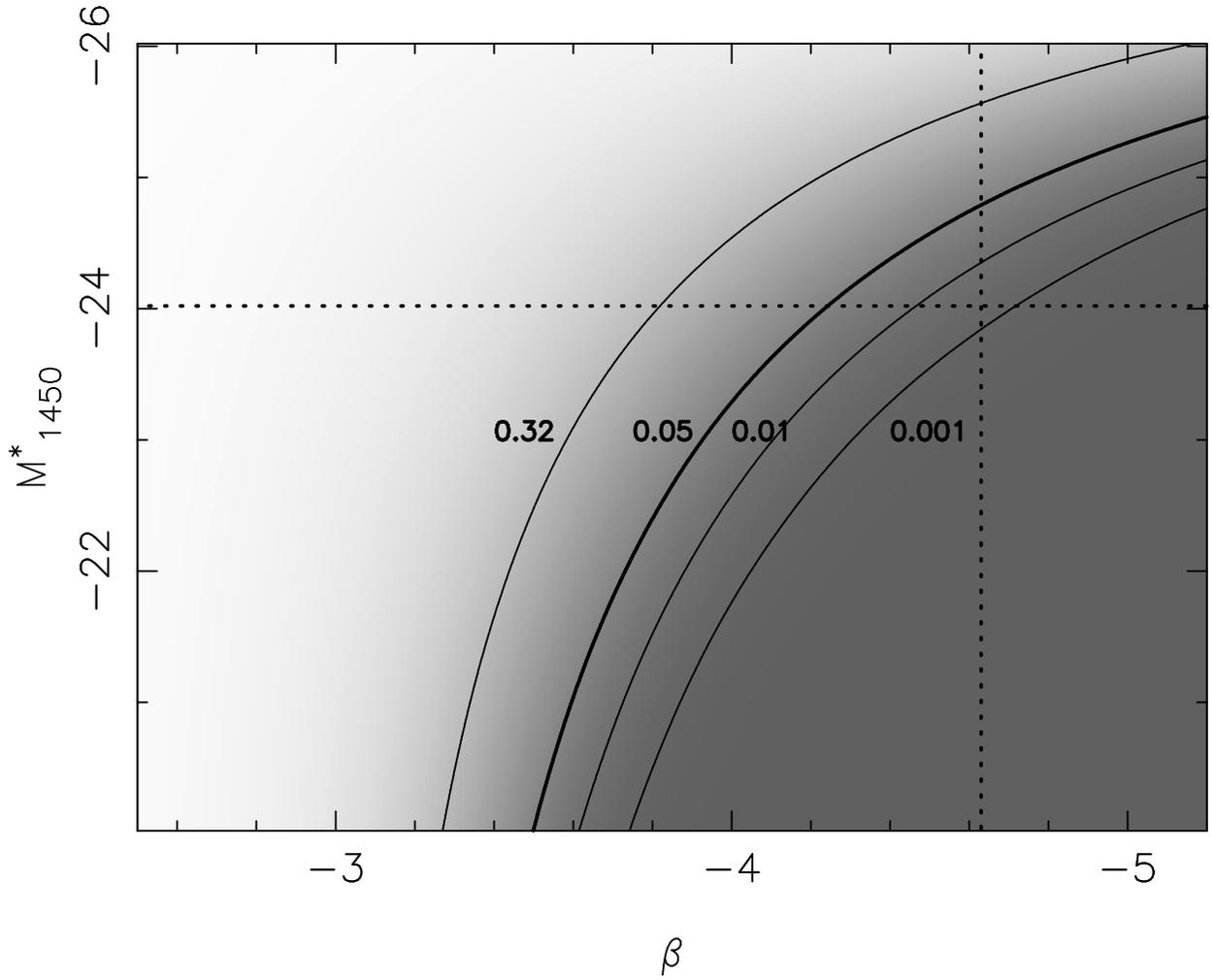}
\caption{Contours of fixed likelihood for no lensing among the four
$z>5.7$ quasars, shown in the two--dimensional parameter space of the
slope and break of the $z\sim 6$ quasar luminosity function.  The
vertical dotted line shows the $3\sigma$ limit on $\beta$ given an
assumed break in the luminosity function of $M_{1450}^\ast=-24.0$ ({\em
horizontal dotted line}).\label{fig:contours}}
\end{figure}


\begin{deluxetable}{lllll}
\tabletypesize{\small}
\tablewidth{0pt}
\tablecaption{\label{tab:tab1}}
\tablehead{
\colhead{SDSSp~J} &
\colhead{Redshift} &
\colhead{SDSS $z$} &
\colhead{HST $z$} &
\colhead{Reference}
}
\startdata
104433.04$-$012502.2 & 5.74 & 19.3 & 19.91 & 1 \\
083643.85+005453.3 & 5.82\tablenotemark{a} & 18.79 & 19.49 & 2 \\
130608.26+035626.3 & 5.99 & 19.46 & 20.23 & 2 \\
103027.10+052455.0 & 6.30\tablenotemark{b} & 20.02 & 20.54 & 2 \\
\enddata
\tablecomments{Column (1) lists the J2000 coordinates, column (2) the redshift, columns (3) and (4) give SDSS and {\em HST} $z$-band magnitudes and column (5) gives the discovery references, which are are (1) \markcite{fwd+00}{Fan} {et~al.} (2000) and (2)
\markcite{fnl+01}{Fan} {et~al.} (2001a).}
\tablenotetext{a}{\markcite{ste+03}{Stern et al.} (2003) find $z=5.77$ from CIII] emission.}
\tablenotetext{b}{Corrected from $z=6.28$ using a fit to templates that
allow for emission line blueshifts \markcite{rvr+02}({Richards} {et~al.} 2002).}
\end{deluxetable}

\begin{deluxetable}{lrrr}
\tabletypesize{\small}
\tablewidth{0pt}
\tablecaption{\label{tab:tab2}}
\tablehead{
\colhead{$z$} &
\colhead{$4$} &
\colhead{$5$} &
\colhead{$6$}
}
\startdata
$M_{1450}$ & -25.48 & -26.05 & -26.51 \\
$M_{1450}^*$ & -24.73 & -24.42 & -24.12 \\
diff & 0.75 & 1.63 & 2.39 \\
$\tau_2$ & 0.21\% & 0.29\% & 0.37\% \\
Prob. & 1.2\% & 3.5\% & 8.2\% \\
\enddata
\tablecomments{The rows are (1) the absolute magnitude at $z=20.2$
mag, (2) $L^*$ from the \markcite{wl02}{Wyithe} \& {Loeb} (2002a) quasar LF, (3) the difference of
the above two, (4) the multiple lensing optical depth (i.e., fraction of lines of sight subject to lensing), and (5) the
lensing optical depth with bias, assuming slope $\beta_h=3.5$.}
\end{deluxetable}

\end{document}